\documentclass[twocolumn,10pt]{article}
\usepackage[a4paper, left=1.6cm, right=1.6cm, top=1.9cm, bottom=2.54cm]{geometry}
\usepackage{times}           
\usepackage{amsmath, amssymb}
\usepackage{graphicx}
\usepackage{caption}
\usepackage{titlesec}
\usepackage{fancyhdr}
\usepackage{hyperref}
\usepackage[table]{xcolor}
\usepackage{cite}
\usepackage{amsmath,amssymb,amsfonts}
\usepackage{algorithmic}
\usepackage{graphicx,color}
\usepackage{textcomp}
\usepackage[table]{xcolor}
\usepackage{cite}
\usepackage{amsmath,amssymb,amsfonts}
\newcommand{\ignore}[1]{ }
\usepackage{algorithmic}    
\usepackage{graphicx}
\usepackage{textcomp}
\usepackage{xcolor}
\usepackage{lmodern}
\usepackage{float}
\usepackage{multirow}
\usepackage{lipsum}
\usepackage{listings}
\usepackage{color}
\usepackage{amsmath}
\usepackage{subcaption}

\definecolor{dkgreen}{rgb}{0,0.6,0}
\definecolor{gray}{rgb}{0.5,0.5,0.5}
\definecolor{mauve}{rgb}{0.58,0,0.82}

\usepackage{adjustbox}
\usepackage{array}

\usepackage[prependcaption, textsize=footnotesize]{todonotes}
\usepackage{todonotes}

\usepackage{pifont}
\newcommand{\cmark}{\ding{51}}%
\newcommand{\xmark}{\ding{55}}%

\usepackage{pifont}
\newcolumntype{P}[1]{>{\centering\arraybackslash}p{#1}}

\newcounter{mysfig}

\renewcommand\themysfig{\theFig.(\alph{mysfig})}
\makeatletter
\newcommand\Scaption[1]{%
\vskip.5\abovecaptionskip
  \sbox\@tempboxa{\small~#1}%
  \ifdim \wd\@tempboxa >\hsize
    \small\themysfig~#1\par
  \else
    \global \@minipagefalse
    \hb@xt@\hsize{\hfil\box\@tempboxa\hfil}%
  \fi
  \vskip\belowcaptionskip}
\makeatother
\usepackage{array}

\usepackage{balance}
\usepackage{hyperref}

\usepackage{courier}

\usepackage{ragged2e}

\usepackage{array} 

\usepackage{tikz,xcolor,hyperref}

\definecolor{lime}{HTML}{A6CE39}
\DeclareRobustCommand{\orcidicon}{%
	\begin{tikzpicture}
	\draw[lime, fill=lime] (0,0) 
	circle [radius=0.16] 
	node[white] {{\fontfamily{qag}\selectfont \tiny ID}};
	\draw[white, fill=white] (-0.0625,0.095) 
	circle [radius=0.007];
	\end{tikzpicture}
	\hspace{-2mm}
}

\foreach \x in {A, ..., Z}{%
	\expandafter\xdef\csname orcid\x\endcsname{\noexpand\href{https://orcid.org/\csname orcidauthor\x\endcsname}{\noexpand\orcidicon}}
}

\def\BibTeX{{\rm B\kern-.05em{\sc i\kern-.025em b}\kern-.08em
    T\kern-.1667em\lower.7ex\hbox{E}\kern-.125emX}}
\AtBeginDocument{\definecolor{ojcolor}{cmyk}{0.93,0.59,0.15,0.02}}
\def\OJlogo{\vspace{-4pt}\hskip-4pt\includegraphics[height=18pt]{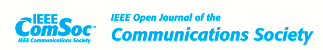}}
\begin{document}

\title{Bridging Subjective and Objective QoE: Operator-Level Aggregation Using LLM-Based Comment Analysis and Network MOS Comparison}

\author{Parsa Hassani Shariat Panahi \orcidB, Amir Hossein Jalilvand \orcidA, and M. Hassan Najafi \orcidC}


\maketitle

\begin{abstract}
This paper introduces a dual-layer framework for network operator-side quality of experience (QoE) assessment that integrates both objective network modeling and subjective user perception extracted from live-streaming platforms. On the objective side, we develop a machine learning model trained on mean opinion scores (MOS) computed via the ITU-T P.1203 reference implementation, allowing accurate prediction of user-perceived video quality using only network parameters such as packet loss, delay, jitter, and throughput--without reliance on video content or client-side instrumentation. On the subjective side, we present a semantic filtering and scoring pipeline that processes user comments from live streams to extract performance-related feedback. A large language model is used to assign scalar MOS scores to filtered comments in a deterministic and reproducible manner.

To support scalable and interpretable analysis, we construct a labeled dataset of 47,894 live-stream comments, of which approximately 34,000 are identified as QoE-relevant through multi-layer semantic filtering. Each comment is enriched with simulated Internet Service Provider attribution and temporally aligned using synthetic timestamps in 5-minute intervals. The resulting dataset enables operator-level aggregation and time-series analysis of user-perceived quality. A delta MOS 
($\Delta \text{MOS}$) metric is proposed to measure each Internet service provider's deviation from platform-wide sentiment, allowing detection of localized degradations even in the absence of direct network telemetry. A controlled outage simulation confirms the framework's effectiveness in identifying service disruptions through comment-based trends alone.

The system provides each operator with its own subjective MOS and the global platform average per interval, enabling real-time interpretation of performance deviations and comparison with objective network-based QoE estimates. The proposed architecture is scalable, privacy-preserving, and infrastructure-agnostic, offering a practical solution for enhancing operator awareness through platform-derived user feedback.
\end{abstract}


\section{Introduction}
\ignore{
\IEEEPARstart{E}{xcellent} user experiences are fundamental in modern communication systems. This requires ensuring seamless interactions, intuitive designs, and dependable systems. Such efforts contribute to increased customer satisfaction and loyalty, enabling businesses to succeed in competitive markets. Understanding and improving user experiences is vital for maintaining a competitive advantage and meeting customer expectations \cite{mine, arellano2024survey}. Over the years, extensive research has been conducted on QoE, resulting in the development of various QoE assessment frameworks \cite{sr-8, sr-11, sr-1, sr-12, sr-20}. For instance, \texttt{YoMoApp} \cite{sr-1, sr-12} evaluates key performance indicators (KPIs) for YouTube video QoE in mobile networks. As described in \cite{sr-8}, a framework using a video server, network simulator, and receiver examines the impact of network parameters on IPTV user QoE. Reference \cite{sr-11} introduces a tool that enhances mobile network performance by integrating subjective user QoE data with objective network parameters. In \cite{sr-20}, a proactive LTE network management framework based on user QoE is proposed, which utilizes ML algorithms to predict QoE parameters using extensive cell-level network performance data.}

The exponential growth of internet-based services and multimedia applications has dramatically increased the demand for reliable, user-oriented quality assessments. As network environments become more dynamic and heterogeneous, ensuring optimal user satisfaction requires going beyond traditional quality of service  metrics. This has led to the emergence of quality of experience (QoE) as a more holistic and user-centric indicator that reflects how users perceive service quality under diverse and often unpredictable conditions \cite{mine, arellano2024survey}.

Traditionally, QoE assessment has relied on either intrusive subjective testing, which is resource-intensive and difficult to scale, or objective estimation methods based solely on static network-level parameters. These approaches often fail to account for the dynamic and context-sensitive nature of human perception, resulting in limited accuracy and generalizability--especially in real-world deployment scenarios.

To overcome these challenges, researchers have increasingly turned to hybrid QoE frameworks that fuse objective network indicators with user-centric feedback and machine learning techniques. For example, \cite{intro-1} proposed a model that integrates perceptual factors such as buffering duration alongside physical-layer metrics like RSRP and SINR, using an artificial neural network to improve prediction accuracy in mobile environments. Expanding on this paradigm, Gao et al. \cite{intro-2} emphasized the integration of contextual and psychological dimensions, including device type, emotional state, and usage environment, into deep learning and Bayesian models--enabling richer and more individualized QoE inference.

Lightweight and scalable implementations of such frameworks have also gained traction. Ref. \cite{intro-3}, for instance, provides cross-application QoE estimation using decision tree models trained on passively collected network and device-level metrics. By crowdsourcing end-user feedback and minimizing overhead via UDP-based probing, this framework offers a real-time, practical solution for mobile network monitoring and experience forecasting. In a similar spirit, \cite{intro-4} proposed a refined approach for fixed wireless access  networks, where a calibrated customer experience index aligns conventional service KPIs with subjective user feedback--yielding a low-complexity but effective model for adaptive QoE evaluation.
%
Crowdsourcing has also proven critical for bridging the gap between subjective perception and objective signals. The user-centric web QoE evaluating framework \cite{intro-5} exemplifies this by combining live subjective scores with automated logging of network metrics during real web usage sessions. The resulting data-driven models demonstrated superior accuracy--about 5\% higher than strong machine learning baselines--highlighting the value of integrating user subjectivity into predictive algorithms.

From a methodological standpoint, Charonyktakis et al. \cite{intro-6} introduced  a modular framework designed to dynamically select the most suitable machine learning algorithm (\textit{e.g.,} support vector regression, artificial neural networks, gaussian naive bayes, decision trees) for a given dataset or user profile. Leveraging nested cross-validation and Bayesian feature selection, this framework adapts to both data characteristics and user heterogeneity. The framework consistently outperforms conventional methods such as PESQ and the E-model across VoIP and video streaming datasets, emphasizing the importance of adaptive, per-user modeling for robust QoE prediction.
To this end, we propose a dual-layer QoE assessment framework that bridges standardized network-side metrics with user-perceived quality extracted from live content platforms. On the network operator side, objective MOS is estimated using an enhanced ITU P.1203-compliant model \cite{mine-ieee} applied to key performance indicators. In parallel, subjective MOS is inferred from user-generated feedback through large language model (LLM) analysis. This integrated approach enables comparative QoE evaluation across operators and platforms.

\section{Contributions}

This research's primary contribution lies in developing a dual-layer QoE assessment framework that facilitates operator-level comparison between subjective and objective quality metrics across mobile and multimedia networks. The framework synthesizes LLM-processed mean opinion scores (MOS) from user-generated content and contrasts them with network parameter-driven MOS predictions generated by our enhanced ITU P.1203-compliant machine learning model.

The principal innovations of this work include:

\begin{itemize}
\item \textbf{Aggregated Subjective QoE Estimation:} We introduce a processing pipeline that systematically filters and evaluates real-world user feedback from live streaming platforms through semantic analysis and LLM integration. The derived quality scores are aggregated according to network operator identification (via IP ranges), establishing a scalable framework for passive subjective QoE assessment.

\item \textbf{Objective QoE Estimation Through ITU-Compliant Modeling:} Building on our prior research, we implement an optimized Random Forest model inspired by ITU P.1203 standards to calculate objective MOS from network key performance indicators (KPIs) including latency, jitter, packet loss, and bitrate. These metrics are aggregated per operator for comparative analysis at the group level.

\item \textbf{Subjective–Objective MOS Discrepancy Analysis:} Through systematic alignment of comment-derived subjective MOS with network-based objective MOS at the operator level, our framework identifies critical QoE mismatches – particularly instances where technical measurements overestimate actual user satisfaction or fail to detect service quality issues.

\item \textbf{Three-Tier QoE Architecture Implementation:} Our solution organizes quality metrics into an Access-Service-Experience hierarchy, building upon established QoE modeling literature \cite{llm-1, j1, llm-3} while maintaining alignment with industry-standard monitoring architectures like ETSI TS 103 294. This stratified approach enables coherent interpretation of both network performance metrics and user quality perceptions through operator-centric aggregation.

\item \textbf{Open Architecture with Operational Scalability:} All core components – including content filtration algorithms, MOS calculation modules, and data aggregation systems – are developed with reproducibility and open-source principles. The privacy-conscious design supports real-time deployment in operational QoE monitoring dashboards while ensuring scalability across network infrastructures.
\end{itemize}

\subsection{Paper Structure}

In §\ref{related-works}, we review related work on QoE estimation and the use of LLMs in subjective quality assessment.
§\ref{proposed-framework} introduces our proposed framework, detailing the semantic filtering pipeline, MOS estimation model, and aggregation methodology.
In §\ref{dataset}, we describe the dataset used in our experiments, including comment preprocessing, labeling, and timestamp simulation.
§\ref{results} presents the results and discusses how operators can interpret QoE from platform-side comment analysis, including $\Delta \text{MOS}$ trends and outage detection.
Finally, §\ref{conclusion} concludes the paper with a summary of contributions and outlines directions for future research.

\section{Related Works}
\label{related-works}
\begin{table*}[]

\caption{
Comparison of recent works on QoE and QoS estimation in multimedia and service platforms. The table summarizes key contributions of each study across dimensions such as the use of subjective and objective inputs, integration of LLMs, ability to produce QoE/QoS outputs, bridging of subjective and objective layers, support for operator-level insights, and real-time applicability. Subj.: subjective input (\textit{e.g.,} comments, speech); Obj.: objective input (\textit{e.g.,} network or video metrics)}
\label{tbl:r-w}
\begin{tabular}
{|P{2cm}|p{3cm}|P{1cm}|P{1cm}|P{1cm}|P{1.5cm}|P{1cm}|P{1.5cm}|P{1.8cm}|}

\hline
\textbf{Refrence}                  & \textbf{Main Idea}                                         & \textbf{Subj. Input} & \textbf{Obj. Input} & \textbf{LLM Usage}    & \textbf{QoE/QoS Output} & \textbf{Bridges Subj.-Obj.} & \textbf{Operator-Level Insight} & \textbf{Real-Time Capability} \\ \hline
Baraković et al. {\cite{f4-2}}       & Early modeling and monitoring of QoE                       & \xmark     & \xmark    & \xmark & \xmark   & \xmark       & \xmark           & \xmark         \\ \hline
Sultan et al. {\cite{f4-3}}          & QoE evaluation in high-bandwidth comms                     & \cmark     & \cmark    & \xmark & \cmark   & \xmark       & \cmark           & \xmark         \\ \hline
Barman et al. {\cite{f4-4}}         & Survey of adaptive streaming QoE models                    & \xmark     & \cmark    & \xmark & \cmark   & \xmark       & \xmark           & \xmark         \\ \hline
Liotou et al. {\cite{f4-5}}         & Caching impact on QoE via ITU model                        & \xmark     & \cmark    & \xmark & \cmark   & \xmark       & \cmark           & \xmark         \\ \hline
Barakabitze et al. {\cite{f4-6}}    & Future-oriented QoE management with ML                     & \cmark     & \xmark    & \xmark & \cmark   & \xmark       & \cmark           & \xmark         \\ \hline
Kougioumtzidis et al. {\cite{j1}} & ML-based prediction of QoE                                 & \xmark     & \cmark    & \xmark & \cmark   & \xmark       & \cmark           & \xmark         \\ \hline
Omar et al. {\cite{f4-7}}           & ML-based QoE prediction using network data                 & \cmark     & \cmark    & \xmark & \cmark   & \xmark       & \cmark           & \xmark         \\ \hline
Panahi et al. {\cite{mine-ieee}}         & ITU P.1203-based MOS from network params                   & \xmark     & \cmark    & \xmark & \cmark   & \xmark       & \cmark           & \cmark         \\ \hline
LLM4Band {\cite{llm-4}}              & Bandwidth estimation via LLM + offline RL                  & \xmark     & \cmark    & \cmark & \xmark   & \xmark       & \xmark           & \cmark         \\ \hline
Rezaee \& Gundavelli {\cite{llm-5}}  & Detect verbal/visual signs of poor QoE in calls            & \cmark     & \cmark    & \cmark & \cmark   & \xmark       & \xmark           & \cmark         \\ \hline
Jha \& Chuppala {\cite{llm-6}}       & LLM QoS serving with RL, QUIC scheduling                   & \xmark     & \cmark    & \cmark & \cmark   & \xmark       & \xmark           & \cmark         \\ \hline
Liu et al. {\cite{llm-7}}            & LLM-enhanced QoS prediction from metadata                  & \xmark     & \cmark    & \cmark & \cmark   & \xmark       & \xmark           & \xmark         \\ \hline
\textbf{This Work}             & Bridging subjective (comments) \& objective QoE using LLMs & \cmark     & \cmark    & \cmark & \cmark   & \cmark       & \cmark           & \cmark         \\ \hline

\end{tabular}

\end{table*}

\label{sec:Background}
As summarized in \autoref{tbl:r-w}, prior studies have addressed various facets of QoE assessment, from foundational models \cite{f4-2} to ITU-aligned resource optimization strategies \cite{f4-3}, and perceptual machine learning approaches \cite{f4-4}. While works like Liotou \textit{et al.} \cite{f4-5} and Barakabitze \textit{et al.} \cite{f4-6} advanced human-centric modeling and ML-based QoE management, they lacked complete implementation pipelines for data acquisition and MOS computation. More recent studies \cite{j1, f4-7} offered subjective data tools and ML-based QoE prediction, yet none fully bridged the gap between data collection, ITU-compliant modeling, and end-to-end QoE estimation.

Complementary research in content-level quality analysis has prioritized computational efficiency through selective processing techniques \cite{rev2, rev3}, a philosophy echoed in our network-oriented framework. By addressing prior limitations and unifying subjective and objective QoE estimation, our approach delivers a complete, operator-scalable solution grounded in standard-based modeling and LLM-driven perception analysis.
In our prior research~\cite{mine-ieee}, we developed a novel network-centric QoE assessment system that overcomes critical gaps in existing methodologies. Distinct from earlier approaches, this framework delivers a complete open-source pipeline for calculating ITU P.1203-aligned  MOS using exclusively network-layer metrics---including latency, jitter, packet loss, throughput, and bitrate---eliminating dependency on video content analysis. The system achieves 97\% parity with the full ITU P.1203 reference model while substantially streamlining data acquisition workflows and computational resource requirements.

\begin{itemize}
    \item \textbf{End-to-End Network Parameterization}: Integration of a Selenium-driven data collection engine that automates HAR file generation, segment-level metadata extraction, and MOS computation via the ITU reference model, bypassing content inspection.
    
    \item \textbf{Machine Learning Optimization}: Deployment of a Random Forest regressor trained on 22,000 annotated network-performance samples, enabling direct MOS prediction from network telemetry with high fidelity (R\textsuperscript{2} = 0.968) across diverse emulated network environments.
    
    \item \textbf{Operational Scalability}: A lightweight architecture supporting real-time monitoring scenarios, coupled with full reproducibility through open-sourced codebases, datasets, and implementation protocols.
\end{itemize}
This work advances beyond predecessors like Sultan~et~al.~\cite{f4-3} and Omar~et~al.~\cite{f4-7}, which---despite incorporating machine learning---lack holistic, production-grade implementations. By contrast, our framework provides telecom operators with an immediately deployable solution for QoE optimization using exclusively network-accessible data, eliminating reliance on proprietary content or user-side instrumentation.

Recent works have begun exploring the use of large language models to enhance network-layer QoE systems. For example, LLM4Band \cite{llm-4} introduces a hybrid framework where an LLM augments offline reinforcement learning for more accurate bandwidth estimation in RTC scenarios. While this approach focuses solely on objective estimation and real-time control, it demonstrates the potential of LLMs to generalize across diverse network traces. In contrast, our current work applies LLMs directly to subjective user feedback (\textit{e.g.,} live-streamed comments), enabling end-to-end estimation of perceived QoE and allowing for a direct comparison between network-based and user-reported MOS at the operator level.
Rezaee and Gundavelli \cite{llm-5} proposed a conceptual system for real-time monitoring and enhancement of multi-user communication using AI agents and LLMs. Their approach involves analyzing audio/video streams during teleconferencing sessions to detect verbal and visual indicators of poor Quality of Experience (\textit{e.g.,} "I can't hear you"), and correlating them with network telemetry to generate a real-time QoE signal.

Jha and Chuppala \cite{llm-6} propose a QoS-aware LLM serving framework that dynamically routes requests between LLMs of varying latency and quality using deep reinforcement learning. While their work focuses on backend serving efficiency--using QUIC stream scheduling and model selection to meet latency guarantees--our work addresses a different facet of QoE: measuring user perception from live-streamed comments and aligning it with network-based MOS. Nonetheless, both works reflect an emerging trend of resource-aware LLM systems optimized for application-level performance and user satisfaction.
Liu et al. \cite{llm-7} proposed the llmQoS model for QoS prediction in service recommendation, using LLMs to extract semantic features from textual descriptions of users and services. These LLM-derived embeddings are combined with user-service interaction history to enhance predictive performance under sparse data conditions. While their focus is on objective metrics like throughput and response time, our work extends the role of LLMs into the subjective domain--processing user-generated comments to infer perceived MOS. This highlights a broader trend in QoS/QoE research toward leveraging LLMs to fill semantic or data gaps in traditional estimation pipelines.

\section{Proposed Framework}
\label{proposed-framework}
\begin{figure*}
    \centering
    \includegraphics[width=0.85\linewidth]{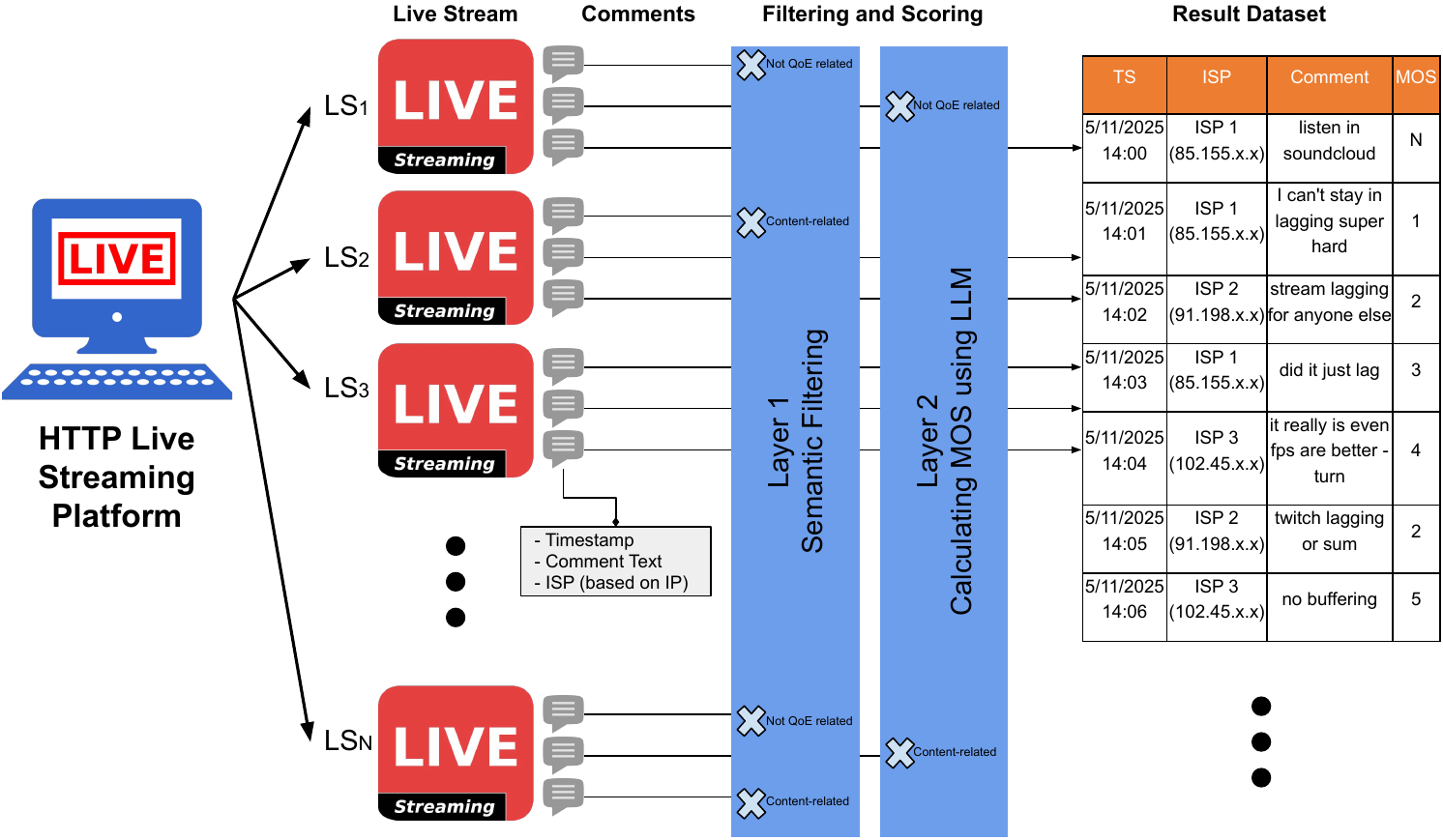}
    \caption{Semantic filtering and LLM-based MOS estimation applied to real-time live stream comments. Each entry is labeled with ISP information and timestamp, enabling subjective QoE analysis across providers.}
    \label{fig:main}
\end{figure*}

In our previous work \cite{mine-ieee}, we introduced a
framework that accurately estimates the MOS using only network-level parameters derived from the ITU-T P.1203 standard, without requiring video content or client-side metrics. Building on that objective-side model, the current work proposes a complementary framework for estimating subjective MOS on the service platform side by analyzing user comments. By semantically filtering and scoring quality-related feedback using a large language model, we compute comment-based subjective MOS values that reflect real user satisfaction \autoref{fig:main}. These subjective scores are then aggregated per operator and aligned with their corresponding objective MOS values. This dual-source approach enables network operators to detect mismatches between perceived and measured quality, improve service diagnosis, and achieve a more accurate and comprehensive understanding of user QoE by bridging the gap between platform-side perception and operator-side delivery metrics.

\subsection{Functional QoE Layer Mapping}


\begin{figure}
    \centering
    \includegraphics[width=0.7\linewidth]{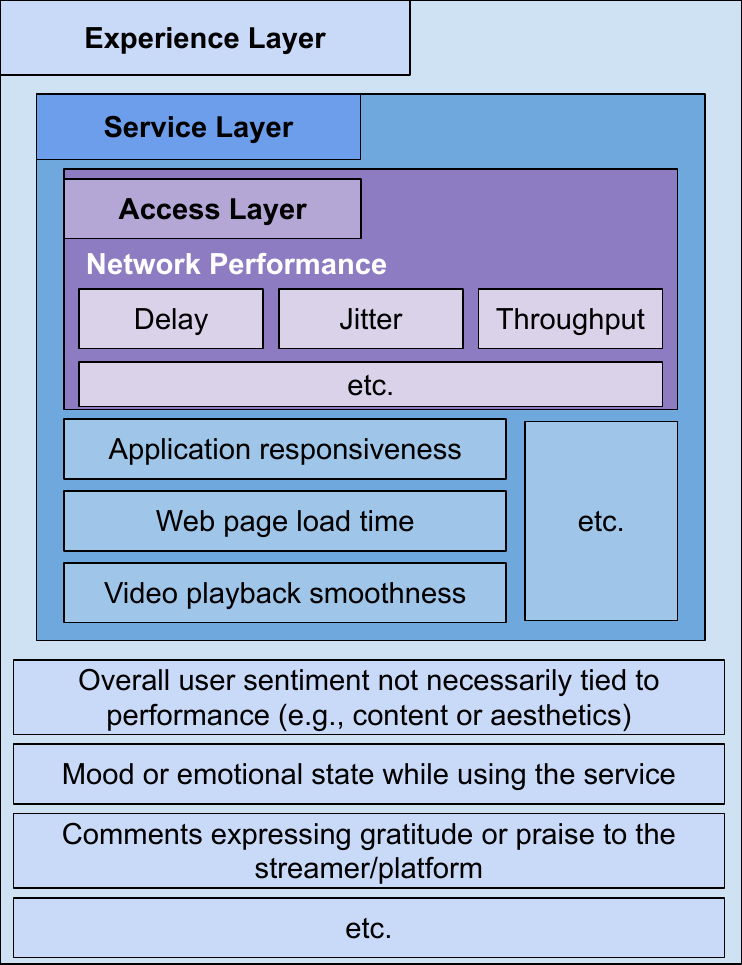}
    \caption{Three-layer QoE model illustrating how objective network metrics (Access Layer), application-level symptoms (Service Layer), and subjective user perceptions (Experience Layer) together form a complete view of user experience.}
    \label{fig:arch}
\end{figure}

To organize the complex range of factors influencing QoE, we adopt a three-level functional model derived from existing QoE-QoS monitoring architectures and modeling literature~\cite{llm-1,j1,llm-3} (see~\autoref{tab:layering}), including ETSI TS~103~294 and the layered decomposition described in~\cite{llm-2}. These models typically distinguish among three key stages: (1) technical delivery, (2) application-level behavior, and (3) user perception. 

As illustrated in~\autoref{fig:arch}, we formalize this structure into a simplified yet operationally relevant architecture consisting of:

\begin{itemize}
    \item \textbf{QoE Part~0 (Access Layer)}: Network-level performance metrics,
    \item \textbf{QoE Part~1 (Service Layer)}: Application observable behavior,
    \item \textbf{QoE Part~2 (Experience Layer)}: User-perceived quality.
\end{itemize}

Each layer corresponds to a distinct stage in the service delivery chain, with progressively decreasing levels of direct observability and controllability by network operators.

\begin{table*}[]
\centering
\caption{
Hierarchical QoE layer model employed in this study. The Access Layer captures raw network performance metrics, while the Service Layer characterizes observable application behavior, influenced by either network conditions or platform-specific implementations. The Experience Layer encompasses user-perceived quality, integrating both lower layers; however, only performance-relevant components are retained within the framework to maintain alignment with operator objectives
}
\begin{tabular}{|p{2cm}|p{3.5cm}|p{3.5cm}|p{3cm}|p{3cm}|}
\hline
QoE Layer & Content Scope                                              & Relation to Other Layers               & Operator Perspective                   & Platform Perspective              \\ \hline
\textbf{Experience Layer (QoE Part 2)}   & User-perceived quality, satisfaction, emotional or content-based feedback                  & Perception of Service \& Access Layers+ content/emotion noise        & - Relevant only when tied to lower-layer issues - Not directly measurable & - Present in comments  - Non-performance content is filtered out \\ \hline
\textbf{Service Layer (QoE Part 1)}      & Observable application-level behaviors (\textit{e.g.,} buffering, lag, responsiveness, load delays) & Can result from Access issues  Or arise from app-side/platform logic & - Not directly measurable - Operationally important                       & - Extracted from filtered user comments                          \\ \hline
\textbf{Access Layer (QoE Part 0)}       & Core network delivery metrics (\textit{e.g.,} delay, jitter, packet loss, throughput)               & Base layer for delivery performance                                  & - Fully measurable - Controllable                                         & - Indirectly inferred from user feedback when problems occur     \\ \hline
\end{tabular}
\label{tab:layering}
\end{table*}

\begin{enumerate}
    \item \textbf{QoE Part 0 (QoS) – Access Layer:}
This layer reflects core network performance metrics that are both measurable and controllable by operators. These include:
\begin{itemize}
    \item Packet loss,
    \item Delay and jitter,
    \item Throughput,
    \item DNS resolution latency,
    \item Handover quality (in mobile scenarios),
    \item Radio coverage.
\end{itemize}

In our framework, these metrics are input to a machine learning model trained on ITU-T P.1203-based labels. The output, an objective MOS, is aggregated by IP range to represent each operator's delivery-layer performance.

\item  \textbf{QoE Part 1 – Service Layer:}
This layer represents observable effects of network performance at the application level, including:
\begin{itemize}
\item Video playback smoothness,
\item Web page load time,
\item Call setup delay,
\item Application responsiveness.
\end{itemize}
These symptoms are not directly measurable from the network side, but are crucial to the end-user experience. In our framework, we infer these indirectly by semantically filtering user comments for expressions of service degradation (\textit{e.g.,} "buffering," "lag," "won't load"). These filtered comments indicate service-layer performance problems, enabling operators to gain insight into disruptions they cannot observe directly.

\item \textbf{QoE Part 2 – Experience Layer:}
This layer captures users' overall satisfaction, emotional response, or subjective quality judgment. Examples include:

\begin{itemize}
\item Video MOS as rated by the user,
\item Voice call clarity ratings,
\item Comments like "this is unwatchable" or "perfect stream today",
\item Overall user sentiment not necessarily tied to performance (\textit{e.g.,} content or aesthetics).
\end{itemize}
Our framework generates subjective MOS scores for filtered comments. However, we strictly include only comments that reflect service-related dissatisfaction or satisfaction, not content preferences or emotional reactions. Thus, while the MOS scores are expressed in subjective form (QoE Part 2), they are semantically bounded to QoE Part 1 phenomena.

\item \textbf{Clarification of Scope:}
To avoid misinterpretation, we clarify that this work does not aim to capture full-spectrum user perception as defined in broader QoE Part 2 or End-User Layer models. Specifically, we exclude:
\begin{itemize}
\item Comments reflecting personal content preferences, such as opinions on streamers, games, or topics,
\item Emotionally expressive remarks that do not relate to technical or performance issues,
\item Aesthetic or visual quality judgments not tied to observable service behavior.
\end{itemize}

All subjective MOS scores in this study are limited to performance-related user feedback that can be semantically mapped to technical degradation or application-level disruptions (\textit{e.g.,} buffering, lag, unresponsiveness). This ensures the comment-derived QoE remains anchored in QoE Part 1 phenomena, supporting meaningful comparison with objective network-based MOS.
\end{enumerate}

\subsection{Comment Collection from Live Streaming Platforms}

Prior studies have explored live stream comments from various perspectives. Some works focus on understanding user intent and providing feedback suggestions using multimodal comment data including text, video, and audio \cite{dg-1}. Others have used time-synced comments to model user behavior patterns \cite{dg-2}, estimate user sentiment \cite{dg-3}, or detect offensive language in Twitch chats using transfer learning techniques \cite{dg-4}. These efforts collectively demonstrate that live comments are a rich source of real-time viewer perception. To support our framework, we leveraged public comment datasets, including the TwitchChat dataset \cite{dg-main-1} to model interaction dynamics, and partially used the Twitch.tv Chat Log Data \cite{dg-main-2} for timestamped records and ISP-level aggregation.

\subsection{Semantic Filtering to Isolate QoE-Relevant Feedback}
\begin{figure}
    \centering
    \includegraphics[width=0.70\linewidth]{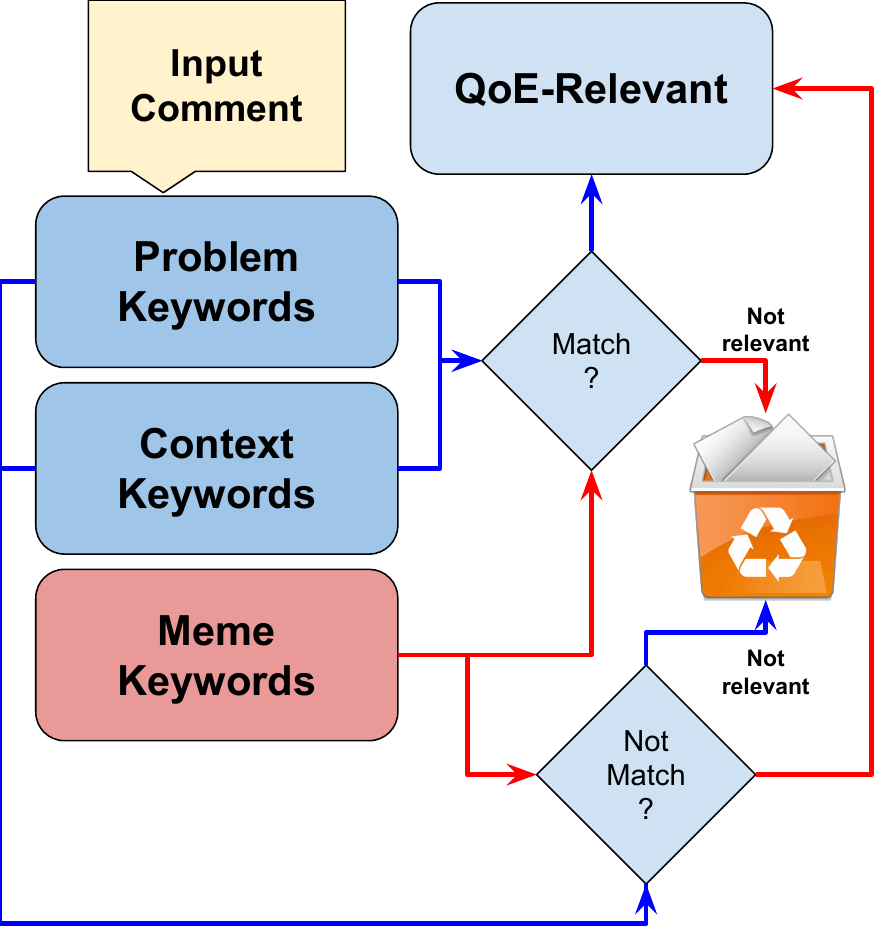}
    \caption{Semantic filtering pipeline for identifying QoE-relevant comments. Each input comment undergoes an initial check against a predefined list of meme keywords; matches are immediately discarded. Non-matching comments are then evaluated for the presence of both problem-related and context-related keywords. Only comments containing at least one problem keyword and one context keyword are classified as QoE-relevant and advanced for further analysis. The figure illustrates the conditional logic: blue arrows denote progression upon meeting criteria, red arrows indicate rejection points, and the green arrow signifies successful QoE-relevant comment identification.}
    \label{fig:semantic-filtering}
\end{figure}
Following the collection of raw live stream comments, we applied a two-layer semantic filtering pipeline designed to extract only those comments indicative of observable delivery and application-level performance issues--i.e., aligned with QoE Part 0 (Access Layer) and QoE Part 1 (Service Layer). This ensures that downstream subjective MOS estimation is grounded in technically meaningful user feedback. Prior work has demonstrated the value of mining unstructured user-generated content for performance insights. For example, \cite{st-2} show that semantic content similarity improves filtering accuracy in recommendation settings, aligning with our use of embedding-based comment filtering. Similarly, \cite{st-3} highlight the importance of integrating subjective QoE feedback into network operations, a goal our approach supports by exposing comment-based MOS estimates to operators (\autoref{fig:semantic-filtering}).

\begin{enumerate}
\item \textbf{Preprocessing and Normalization:}
Each comment undergoes a text cleaning process prior to analysis. We apply a series of transformations to reduce noise and prepare the text for semantic evaluation:
\begin{itemize}
\item Lowercasing for uniformity,
\item 
Character collapse: repeated characters (\textit{e.g.,} "laaag") are collapsed to reduce expressive exaggeration ("lag"),
\item Special character removal to eliminate emojis, punctuation, and Twitch emotes,
\item Whitespace normalization,
Typo corrections: domain-specific substitutions (\textit{e.g.,} "scuffedd" → "scuffed").
\end{itemize}
This normalization reduces lexical variability and strengthens both exact-match and embedding-based feature extraction.
\item \textbf{Keyword and Contextual Match Filtering:}
We construct three keyword groups:
\begin{itemize}
\item Problem keywords: Based on a curated set of quality-related expressions such as "buffer", "lagging", "audio delay", and "disconnect", we define a list of problem keywords that reflect common service or delivery issues described by users.
\item Context keywords: Domain phrases suggesting relevance to video streaming or network (\textit{e.g.,} "stream", "video", "network", "ping", "dns", "connection")
\item Meme keywords: Non-informative platform-specific expressions (\textit{e.g.,} "weirdchamp", "kekw", "monkaw") that indicate off-topic humor or sarcasm.
\end{itemize}

For a comment to be considered QoE-relevant, it must:
\begin{itemize}
\item Match at least one problem keyword,
\item Match at least one context keyword,
\item Not match any meme keyword.
\end{itemize}
These rules filter out irrelevant or aesthetic commentary while preserving performance-related expressions.

\item \textbf{Semantic Similarity Scoring with Anchors:}

We use the all-MiniLM-L6-v2 model from SentenceTransformers%
, a lightweight transformer-based encoder optimized for semantic similarity \cite{st-5}. This model is built on MiniLM architecture and fine-tuned on a large corpus of sentence pairs, producing 384-dimensional embeddings that efficiently capture the semantic meaning of short texts like user comments \cite{st-4}.
To evaluate relevance, cosine similarity is computed between each comment's embedding and a set of predefined anchor embeddings. Cosine similarity measures the cosine of the angle between two vectors in high-dimensional space--values closer to 1 indicate strong semantic alignment. This method allows us to select the most QoE-relevant comments for further scoring and aggregation \cite{st-4}.

Each candidate comment is encoded in the same embedding space and compared to anchor vectors using cosine similarity. The maximum similarity score across all anchors is retained for each comment. Two thresholds are applied:
To ensure robust filtering across diverse comment lengths, we apply length-sensitive similarity thresholds. For short comments (fewer than \texttt{MIN\_WORDS} = 5), a higher threshold of \texttt{SHORT\_TEXT\_SIM\_THRESHOLD} = 0.40 is used. This is because short comments often lack sufficient context, making them more susceptible to false positives in embedding similarity. For longer comments, we apply a lower threshold of \texttt{SIM\_THRESHOLD} = 0.28, as the added semantic richness provides more reliable embedding signals. This thresholding strategy helps balance precision and recall, ensuring only truly QoE-relevant comments are retained.
These thresholds balance recall and precision, filtering out ambiguous or generic expressions while retaining those semantically close to anchor complaints.

\item \textbf{Batched Evaluation and Final Filtering:} The \texttt{full\_optimized\_filter\_batch} function applies the entire filtering pipeline across all comments in a vectorized and batched manner. For each comment, it first cleans and normalizes the text, checks for the presence of meme, context, and problem keywords, and computes semantic similarity with anchor phrases. Based on the word count and similarity threshold, each comment is either retained or discarded. The function returns a boolean mask identifying comments that are considered QoE-relevant:

\begin{itemize}
\item Token length is evaluated,
\item Boolean flags are set for context, problem, and meme match,
\item Cosine similarity is computed,
\item Thresholding is applied based on token length.
\end{itemize}

Only comments satisfying all filtering criteria are labeled as \texttt{is\_loose\_qoe\_candidate = True}. These are then exported to a CSV file (\texttt{loose\_qoe\_candidates\_cleaned.csv}) for downstream processing in the MOS scoring module.
\end{enumerate}
This semantic filtering layer acts as a high-precision selector for QoE-relevant feedback from a noisy, open-domain comment stream. It enables the platform to isolate only those comments which likely reflect service impairments or delivery issues relevant to operators, thus aligning subjective MOS scoring with technically actionable insights.

\subsection{Subjective MOS Estimation Using Large Language Models}
\begin{figure}
	\centering
	\includegraphics[trim={0cm 0cm 0cm 0cm},clip,width=0.21\textwidth,angle=90]{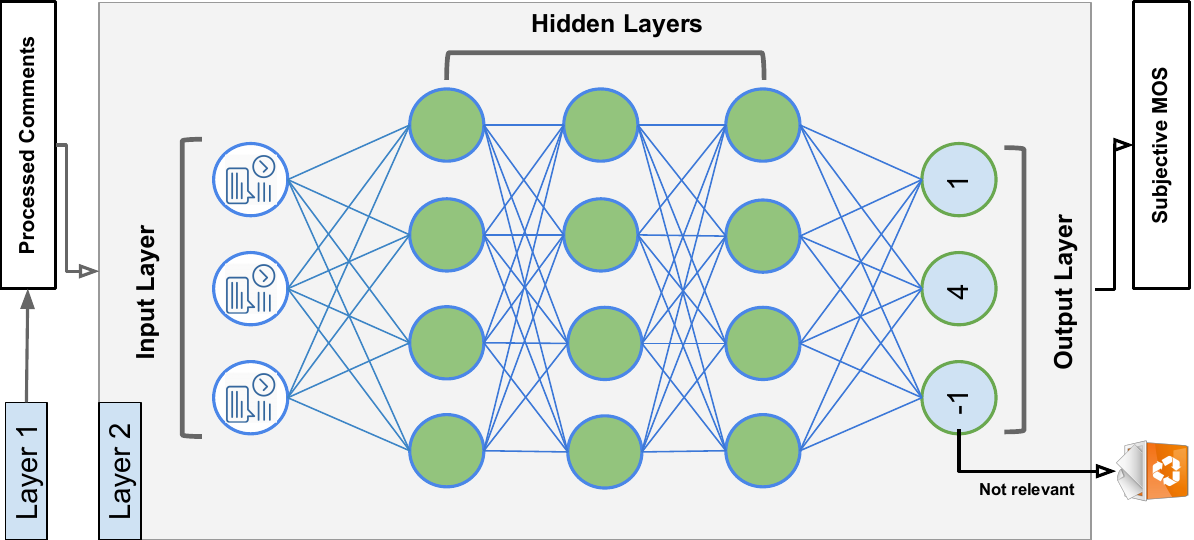}
	\caption{ Subjective MOS estimation using a LLM. After semantic filtering in Layer 1, the processed comments are passed to Layer 2, where a transformer-based LLM interprets each input through multiple hidden layers. These layers apply masked self-attention and feedforward mechanisms to capture nuanced quality-related semantics. The model outputs a scalar value from the output layer, corresponding to a predicted MOS score (\textit{e.g.,} 1–5) or –1 for comments deemed not relevant. Non-relevant predictions are discarded, while valid subjective MOS scores are collected for downstream QoE analysis.}

\label{fig:llm}
\end{figure}
Recent research has demonstrated the potential of LLMs for subjective quality evaluation across various modalities. For instance, \cite{mos-llm-1} showed that auditory LLMs can accurately predict the MOS, perform A/B testing, and generate natural language quality explanations for speech content, with high correlation to human judgments. Similarly, \cite{mos-llm-2} evaluated the ability of instruction-tuned models such as GPT, Gemini Pro, and Claude2 to capture and express subjective assessments. While results vary across models, the study confirms that subjective interpretation is feasible with appropriate prompting. Broader reviews of LLM alignment, such as \cite{mos-llm-3}, highlight the contextual nature of subjective feedback and the importance of feedback-driven tuning to align outputs with human preferences--an idea closely aligned with our approach. Additionally, \cite{mos-llm-4} provide a comprehensive taxonomy of uncertainty estimation techniques in LLMs, emphasizing the challenges of overconfidence in scalar predictions.

After filtering QoE-relevant comments, we apply a transformer-based LLM to estimate subjective MOS based on the semantic content of each comment. As shown in \autoref{fig:llm} this step enables quantification of user-perceived service quality using natural language inputs, bridging the gap between qualitative expressions and numerical evaluation.

\begin{enumerate}
\item \textbf{Model Characteristics and Technical Specifications:}
The language model used for subjective MOS estimation is a multilingual, multimodal, instruction-tuned transformer released in 2024. It is built on a decoder-only architecture and fine-tuned for high-precision, task-specific generation, enabling robust handling of structured instructions. The model supports text, image, and audio modalities, though in this work, only its text-processing capabilities are utilized for evaluating user comments.

It is designed for multilingual deployment, covering more than 50 languages and over 97\% of global speakers. With a context window of up to 128,000 tokens, the model allows flexible prompt design, suitable for contextualized feedback evaluation. Its inference performance is optimized for low latency and high throughput, making it cost-effective and scalable for large-volume subjective analysis tasks.

The model's knowledge cutoff is October 2023, and no external retrieval or real-time access is enabled during inference, ensuring responses are grounded solely in prompt content. To ensure reproducibility and scoring stability, we configure the model in deterministic mode (temperature = 0.0). Though the exact parameter count remains undisclosed, the model has demonstrated state-of-the-art performance on subjective interpretation tasks, as evidenced by its use in recent studies on speech MOS prediction \cite{mos-llm-1}, subjective preference extraction \cite{mos-llm-2}, and feedback-aligned generation \cite{mos-llm-3}.

In line with observations from recent work on LLM uncertainty \cite{mos-llm-4}, our pipeline adopts a simplified verbalization-based scoring strategy, which elicits scalar responses directly through structured prompts. These properties make the model highly suitable for semantic interpretation and subjective scalar scoring, particularly for assigning MOS values based on natural user feedback in streaming contexts.

\item \textbf{Model Architecture:}
The LLM, which is used, is built on a decoder-only transformer architecture, consistent with the design principles introduced in earlier models such as GPT-3 \cite{llm-intro-2}, but with several architectural and efficiency enhancements. In the decoder-only setup, input tokens are processed sequentially using causal (autoregressive) self-attention, where each token attends only to previous tokens in the sequence. 

This left-to-right structure is critical for coherent and fluent language generation. Text input is first tokenized using a byte-level tokenizer, which improves upon previous tokenization schemes by reducing token counts for non-Latin scripts and compressing multilingual input more effectively \cite{llm-intro-1}. Each token is mapped to a dense embedding vector, and positional encodings are added to preserve sequence order. These embeddings are then passed through a deep stack of transformer decoder layers, each consisting of a LayerNorm, a masked multi-head self-attention mechanism, and a feed forward network (FFN), typically implemented as a two-layer MLP with non-linear activation functions such as GELU. Residual connections follow both the attention and FFN sublayers to support gradient flow and model depth. The LLM incorporates optimized attention mechanisms and parallelized computation strategies to support long sequences--up to 128,000 tokens--while maintaining memory and compute efficiency \cite{llm-intro-1}.

Although OpenAI has not disclosed the model's exact internal optimizations (\textit{e.g.,} rotary embeddings or attention compression), its design allows for high-throughput, low-latency inference. In our implementation, the LLM is used with deterministic decoding (temperature = 0.0) and without sampling constraints (\texttt{top\_k}, \texttt{top\_p}), enabling consistent and reproducible scalar outputs. We operate the model in single-turn inference mode, without requiring session memory across prompts, making it ideal for one-shot subjective scoring tasks such as comment-level MOS estimation. The final hidden state is projected into vocabulary space via a linear output head, producing a probability distribution from which the next token is selected or, in our case, a numeric value is extracted directly from the model's output. This configuration leverages the LLM's capabilities in structured reasoning and scalar output generation for comment-driven QoE scoring in a stable, cost-efficient manner.

\item 
\textbf{Prompt Design and Instructional Framing:}
Each user comment is embedded into a pre-defined prompt designed to elicit a discrete satisfaction score. The prompt introduces the comment in the context of using an online service (such as streaming, gaming, or browsing), followed by a description of a 5-point MOS rating scale:
\begin{enumerate}
\item Very Dissatisfied: Severe QoE problems; unusable experience (\textit{e.g.,} constant lags, freezes, or crashes),
\item Dissatisfied: Major recurring issues; frequent buffering or disconnects,
\item Neutral: Minor or occasional QoE disruptions,
\item Satisfied: Good service with negligible issues,
\item Very Satisfied: Perfect experience; no noticeable performance problems.
\end{enumerate}

To maintain data integrity, the prompt includes an explicit fallback condition: if the comment is completely unrelated to user experience or quality, the model is instructed to return –1. This prevents irrelevant or misclassified inputs from distorting subjective MOS distributions.

\item \textbf{Inference Configuration and Execution:}
Each comment is scored using the LLM in isolation via a single prompt-response interaction. We configure the model for deterministic output by setting \texttt{temperature = 0.0}, removing sampling randomness, and ensuring stable and reproducible scores. The interaction is strictly structured, and the model is instructed to return only a single numeric value, eliminating extraneous text.

A dedicated scoring function wraps this logic. It:

\begin{itemize}
\item Embeds the comment into the pre-defined prompt,
\item Sends it to the LLM for evaluation,
\item Parses and validates the output,
\item Handles potential parsing errors gracefully,
\item Logs progress to the console for each step in the loop.
\end{itemize}

A one-second delay is introduced between each scoring operation to respect model rate limits and reduce the risk of throttling.

\item \textbf{Postprocessing and Dataset Generation:}
Each scored comment is stored in a structured dictionary containing:
\begin{itemize}
\item The original text,
\item The assigned MOS value (1–5 or –1 for unrelated).
\end{itemize}

These records are aggregated into a dataset for downstream analysis. The final output represents a comment-level QoE scoring dataset.
\end{enumerate}

\subsection{Simulation of Timestamps and ISP Labeling for Aggregation}
To enable operator-level insights and temporal analysis from subjective QoE scores, we simulate realistic conditions by assigning each comment to an Internet service provider (ISP) and generating synthetic timestamps. This metadata allows us to aggregate MOS over time and across simulated service providers, enabling the detection of performance trends, degradations, or anomalies.

\begin{enumerate}
\item  \textbf{Filtering Valid MOS Data:} 
    After excluding all entries marked as "–1" by the LLM (indicating non-relevant comments), we retain approximately "40,000" valid comment-level MOS records. These represent user perceptions that are semantically grounded in QoE-related issues and ready for aggregation and analysis.

\item \textbf{ISP Assignment and Timestamp Generation:}
    To simulate a diverse network environment, each comment is randomly assigned to one of three hypothetical ISPs: ISP1, ISP2, or ISP3. This step reflects how different user comments might be tied to different service providers in a real-world deployment. The randomness simulates a reasonable distribution of users across ISPs and helps test the system's ability to identify ISP-specific anomalies.

In parallel, we simulate timestamp metadata by assigning a synthetic timestamp to each comment using 3-second intervals starting from a fixed base time (\textit{e.g.,} 2024-01-01 12:00:00). This technique produces a chronologically spaced stream of feedback, as if comments were received continuously during a live session or over a measurement period.
For example:
\begin{itemize}
\item Comment 0 → Timestamp = 2024-01-01 12:00:00, ISP = ISP2
\item Comment 1 → Timestamp = 2024-01-01 12:00:03, ISP = ISP1
\item Comment 2 → Timestamp = 2024-01-01 12:00:06, ISP = ISP3
\end{itemize}

\item \textbf{Time Windowing and Aggregation:}
To support meaningful analysis over time, we group the simulated timestamps into 5-minute windows using time flooring (\texttt{timestamp.dt.floor('5min')}). Within each time window and for each ISP, we compute two key statistics:

\begin{itemize}
    \item Average MOS -- representing the mean perceived service quality
    \item Comment Count -- indicating volume or user activity
\end{itemize}

\begin{table}[]
\centering
\caption{Example of aggregated subjective MOS data grouped by 5-minute time windows and ISP. Each row represents the average user-perceived service quality and number of valid QoE-related comments recorded within a specific time interval for a given provider. This structure enables temporal and cross-operator QoE analysis.}
\label{tab:res-mos-example}
\begin{tabular}{|l|l|l|l|}

\hline
\multicolumn{1}{|c|}{\textbf{Time Window}} & \multicolumn{1}{c|}{\textbf{ISP}} & \multicolumn{1}{c|}{\textbf{Comment Count}} & \multicolumn{1}{c|}{\textbf{Avg MOS}} \\ \hline
12:00–12:05                                & ISP1                              & 420                                         & 3.64                                  \\ \hline
12:00–12:05                                & ISP2                              & 415                                         & 2.81                                  \\ \hline
12:00–12:05                                & ISP3                              & 402                                         & 4.12                                  \\ \hline
12:05–12:10                                & ISP1                              & 435                                         & 3.58                                  \\ \hline
...                                        & ...                               & ...                                         & ...                                   \\ \hline
\end{tabular}
\end{table}
\end{enumerate}
\autoref{tab:res-mos-example} results in a windowed table of per-ISP QoE over time
This format enables direct comparison of service quality across ISPs and across time, supporting both real-time dashboards and offline analysis.

\section{Dataset}
\label{dataset}
\subsection{Dataset Composition}
The final dataset used in this study consists of 47{,}894 comment-level entries, each associated with a subjective MOS generated by an LLM. These entries were collected and processed following the semantic filtering and scoring pipeline described in earlier sections.
Each record contains the original user comment and its corresponding MOS score, which ranges from 1 to 5 for quality-relevant feedback, and --1 for comments deemed not QoE-relevant. Although the primary analysis in this paper focuses on the 33{,}770 valid entries with scores in the [1--5] range, the full dataset---including the 14{,}124 comments labeled --1---is preserved to support future research. These excluded comments are valuable for tasks such as training classification models to distinguish QoE-related and unrelated feedback, or building LLMs capable of predicting the presence or absence of perceived quality feedback in unstructured user language. The retained 33{,}770 comment-MOS pairs form the basis of all temporal, operator-level, and delta-based analyses in subsequent sections. This design ensures both data fidelity for scoring and flexibility for future machine learning experimentation.

\subsection{Subjective Score Distribution}
\begin{figure}
    \centering
    \includegraphics[width=0.95\linewidth]{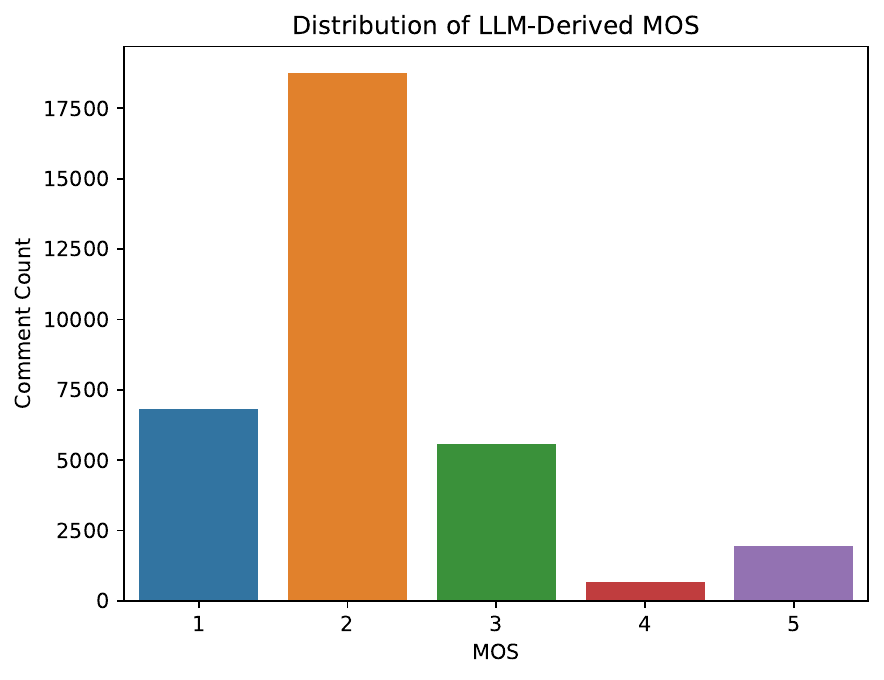}
    \caption{Distribution of MOS derived from LLM evaluations of QoE-related user comments. The distribution is heavily skewed toward lower values, with scores of 1 and 2 dominating. This reflects typical user behavior on live-streaming platforms, where negative experiences prompt explicit feedback, while positive or neutral ones are often unreported or ambiguously expressed.}
    \label{fig:llm-qoe-dist}
\end{figure}

After semantic filtering and LLM-based scoring, the valid portion of the dataset contains 33{,}770 comments with subjective MOS scores ranging from 1 (Very Dissatisfied) to 5 (Very Satisfied). These scores represent the language model's interpretation of user satisfaction based solely on textual feedback.

A distribution analysis reveals that the majority of scores are heavily skewed toward the lower end of the scale, particularly MOS values of 2 and 1. This is an expected and rational pattern in open-user platforms such as live streaming services: users are more likely to vocalize dissatisfaction (\textit{e.g.,} lag, buffering, stuttering) than positive experiences. Positive feedback tends to be rare, short, and ambiguous, while negative feedback is often explicit and tied to technical issues.
Distribution Insights:
\begin{itemize}
\item MOS 2 is the most common label, suggesting widespread expression of service impairments that disrupt but do not fully block usability,
\item MOS 1 appears frequently, corresponding to severe user complaints, often describing unwatchable or broken experiences,
\item MOS 3 appears moderately, often representing minor buffering, momentary lags, or ambiguous tone,
\item MOS 4–5 are the least frequent, aligning with occasional praise or satisfaction that was strong enough to be semantically recognized as quality-related.
\end{itemize}

The bar chart shown in \autoref{fig:llm-qoe-dist} visualizes the comment frequency per MOS level, clearly illustrating the class imbalance. This skew is useful in training or evaluating models under realistic user feedback conditions, where negative sentiment dominates explicitly stated commentary.
This distributional bias also informs future data augmentation strategies or loss function adjustments if the dataset is to be used for supervised learning tasks. Overall, the distribution affirms the practical assumption that subjective QoE signal is louder when quality fails, and more diffuse when performance is acceptable.

\subsection{Simulated Metadata: ISP and Timestamps}
To enable time-series analysis and provider-level aggregation of subjective QoE, each valid user comment in the filtered dataset was enriched with two simulated metadata attributes: an ISP label and a timestamp. This augmentation transforms individual, isolated user feedback entries into temporally and spatially contextualized observations, facilitating longitudinal trend analysis and comparative evaluation across providers. Prior to metadata generation, all comments labeled as –1 by the large language model--indicating irrelevance to service quality--were excluded. The resulting subset of 33,770 valid entries, each associated with a MOS score from 1 to 5, forms the working dataset for simulation and analysis.

To reflect a multi-operator environment, each comment was randomly assigned to one of three synthetic ISPs: ISP1, ISP2, or ISP3. A fixed random seed was applied to ensure reproducibility of provider assignment across experimental runs. Although these assignments are synthetic, they model the operational context in which users are distributed across multiple access networks and service providers.
Simultaneously, we simulated the temporal aspect of feedback by assigning synthetic timestamps to each comment. Starting from a base time of January 1, 2024, at 12:00:00, timestamps were spaced at uniform 3-second intervals, creating a chronologically ordered stream of feedback. This interval approximates a plausible comment frequency during high-traffic live streaming sessions. Each timestamp was then floored to a 5-minute window to support aggregation at a coarser temporal resolution. These time windows serve as the unit of analysis for computing provider-level and global statistics in later sections.

The resulting dataset (as shown in \autoref{tab:final-dataset}) structure includes five fields per record: the user's original comment (\texttt{original\_comment}), the associated subjective MOS score (\texttt{commen\_mos}), the simulated service provider (ISP), the synthetic timestamp (timestamp), and the corresponding aggregation window (\texttt{time\_window}). This metadata-enriched format allows for scalable, window-based analysis of service quality as experienced by users across providers and time.

\begin{table}[]
\centering
\caption{Example entries from the enriched dataset used for subjective QoE analysis. Each record includes a user comment filtered for QoE relevance, a language model–derived MOS score (ranging from 1 to 5), and simulated metadata: a service provider label (ISP1–ISP3), a synthetic timestamp generated at 3-second intervals, and a 5-minute aggregation window used for temporal binning and operator-level analysis.}
\label{tab:final-dataset}
\begin{tabular}{|l|p{2cm}|p{3cm}|l|}

\hline
\multicolumn{1}{|c|}{\textbf{ISP}} & \multicolumn{1}{c|}{\textbf{Timestamp}} & \multicolumn{1}{c|}{\textbf{Original Comment}}    & \multicolumn{1}{c|}{\textbf{MOS}} \\ \hline
ISP1                               & 2024-01-01 12:00:00                     & no buffering                                      & 5.0                               \\ \hline
ISP2                               & 2024-01-01 12:00:03                     & twitch lagging or sum                             & 2.0                               \\ \hline
ISP1                               & 2024-01-01 12:00:06                     & anyone else lose audio at the same time as the dc & 2.0                               \\ \hline
...                                & ...                                     & ...                                               & ...                               \\ \hline
\end{tabular}
\end{table}

\subsection{Temporal Aggregation Format}
To facilitate scalable analysis and time-aware trend detection, all enriched records in the dataset were aggregated into discrete temporal intervals. Each comment's simulated timestamp was binned into a 5-minute window using floor rounding (\texttt{timestamp.dt.floor('5min')}). This transformation allows for smoothing out individual fluctuations in user sentiment and enables operators to analyze trends over short, operationally relevant time frames.
For each unique combination of time window and ISP, we compute two core metrics:

\begin{itemize}
\item Average MOS: The mean subjective score of all comments received from a given ISP during a particular time window.
\item Comment count: The total number of QoE-relevant comments within that window.
\end{itemize}

These aggregated statistics are stored in a derived dataset, indexed by both \texttt{time\_window} and ISP. This structure supports a wide range of time-series operations, including global MOS tracking, comparative provider analysis, and detection of sudden shifts in user-reported service quality.

Formally, for each provider 
$P$ and time window $T$, the average MOS is defined as:
\begin{equation}
\label{formula:avg-mos}
    \text{Avg}_{\text{MOS}_{P,T}} = \frac{1}{N_{P,T}} \sum^{N_{P,T}}_{i=1}{\text{MOS}_i}
\end{equation}
where  $N_{P,T}$ is the number of comments from provider $P$ in time window $T$, and  MOS$_i$ is the LLM-assigned score for comment $i$.
\begin{figure*}
    \centering
        \includegraphics[width=0.95\linewidth]{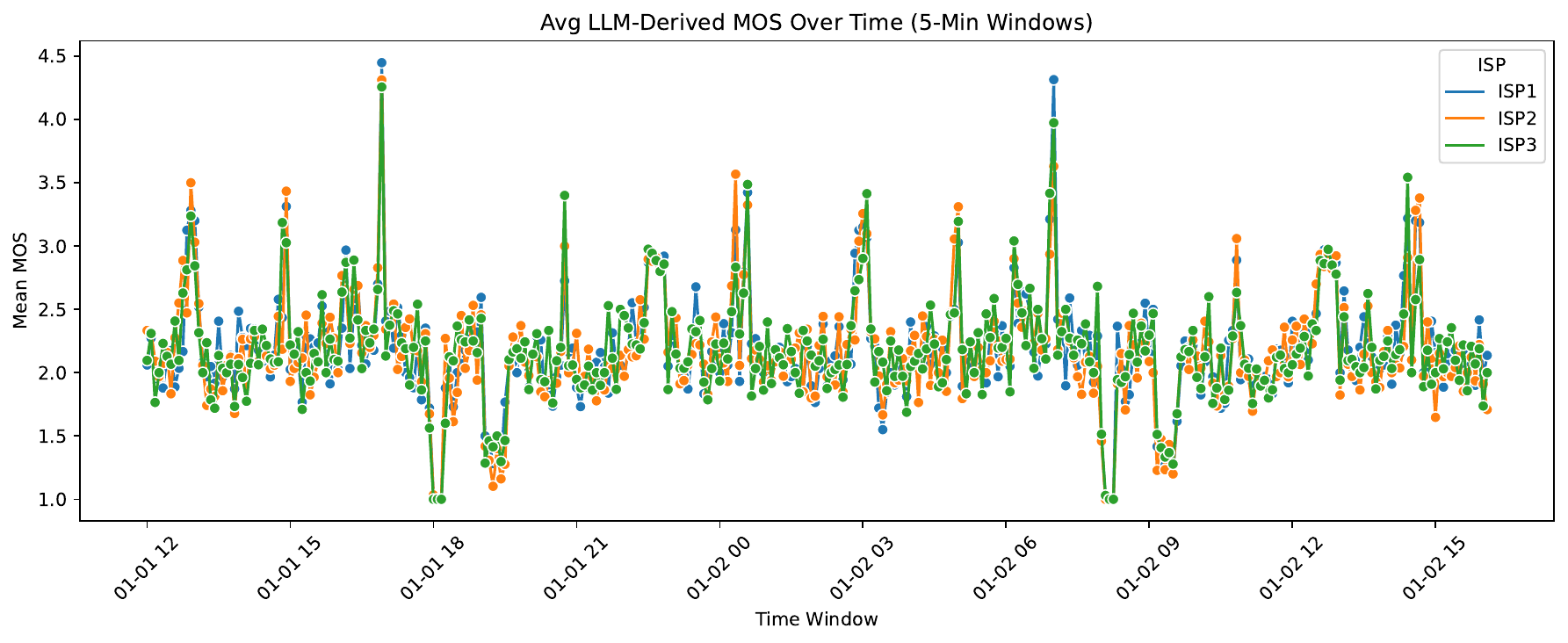}
    \caption{Average MOS over time, aggregated in 5-minute windows and grouped by simulated ISPs (ISP1--ISP3). The values reflect user-perceived quality inferred from comment-level analysis, not direct network metrics. This highlights how platform-side experiences influence perceived quality, enabling comparison of user sentiment across providers and revealing potential service degradations or inconsistencies over time.}
\label{fig:avg-5min}
\end{figure*}

\begin{figure*}
    \centering
    \includegraphics[width=0.95\linewidth]{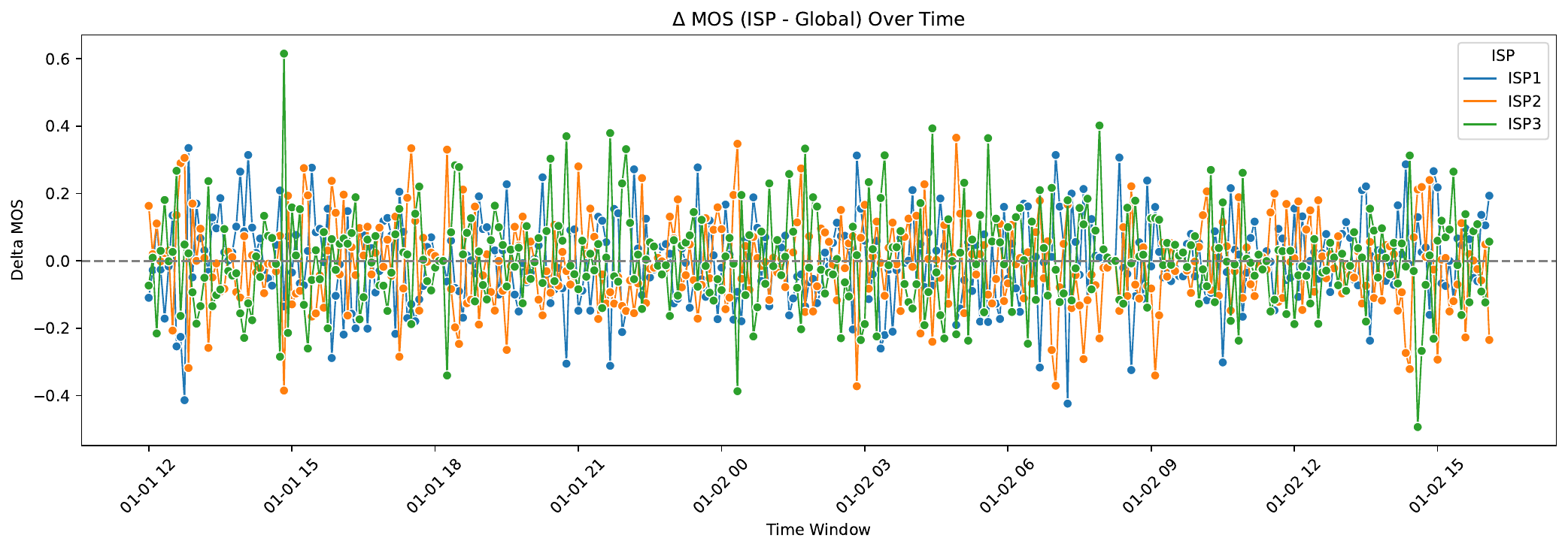}
    \caption{Temporal analysis of $\Delta$ MOS, calculated as the difference between each ISP's average MOS and the global platform-wide MOS in each 5-minute window. Values above 0 indicate higher-than-average user satisfaction for that ISP in a given window, while values below 0 reflect lower relative performance. This view captures subtle service disparities between providers and highlights temporal fluctuations in user experience based on platform-derived subjective feedback, rather than direct network measurements.}
\label{fig:delta-mos}
\end{figure*}

This aggregation approach used in \autoref{formula:avg-mos}, enables flexible downstream analytics, such as identifying degraded intervals, measuring user-perceived service recovery over time, and benchmarking cross-ISP disparities. It also serves as the basis for visualizations and delta analysis presented in the \autoref{results}.

\autoref{fig:avg-5min} illustrates an example of average MOS values over time, aggregated into 5-minute windows and grouped by ISP. The curves reveal temporal patterns of perceived service quality and highlight variability across simulated providers. Although network-side metrics are not directly available, the figure reflects user-perceived performance at the platform level, capturing how streaming conditions manifest in viewer sentiment.
To complement this view, \autoref{fig:delta-mos} presents the $\Delta$MOS the difference between each ISP's average MOS and the global average in the same time window. This relative comparison makes it possible to isolate deviations from baseline satisfaction levels and identify short-term service degradations or improvements. Together, these plots form the foundation for detecting anomalies, evaluating platform impact, and benchmarking simulated ISPs based on platform-side QoE commentary.

\section{Results and Operator-Side QoE Interpretation}
\label{results}
This section analyzes the subjective QoE data generated by our platform-side comment processing pipeline. We examine how aggregated MOS -both globally and per ISP--can inform network operators about user-perceived quality. Through $\Delta \text{MOS}$ analysis and a simulated outage scenario, we show how this framework enables detection of localized degradations and supports real-time QoE monitoring, even without direct access to network-layer metrics.
\subsection{Modeling Assumptions}

To isolate the impact of comment semantics on perceived quality, we assume equal user bases across ISPs and constant comment volume over time. These simplifications help highlight MOS trends from feedback alone, independent of real-world user or traffic variations.
\begin{itemize}
\item \textbf{Equal ISP user:} We assume that all ISPs involved in the simulation serve an equal number of users on the platform. This abstraction enables us to interpret differences in average MOS scores solely as indicators of service quality, without being influenced by disparities in user populations or the relative fraction of QoE-relevant comments per ISP. By neutralizing user volume as a variable, the analysis focuses solely on the semantic content of comments.

\item 
\textbf{Uniform comment volume across time:} It is also assumed that the volume of QoE-related comments remains consistent across all 5-minute time windows throughout the day. In practice, comment activity tends to be higher during peak hours (\textit{e.g.,} 8 PM) and lower during off-peak periods (\textit{e.g.,} 2 AM). However, we intentionally disregard these natural fluctuations to simplify time-series analysis and eliminate the need for temporal normalization.

\end{itemize}

While our assumptions simplify the analysis by treating ISPs as having equal user bases and uniform comment volumes, they abstract away real-world dynamics, such as varying user distributions and temporal comment spikes. In practice, platforms such as Twitch or YouTube have access to detailed user and traffic data, allowing them to compute more accurate, weighted MOS and $\Delta\text{MOS}$ values. These platforms can enhance the framework by adjusting for user volume and engagement trends across ISPs and timeframes.

\subsection{Platform-Wide MOS Calculation and Validation}
A core objective of our platform-side analysis is to provide each network operator with a reliable and interpretable reference for comparing their users' perceived  QoE against broader service-level sentiment. To this end, we compute a platform-wide MOS for each time window, serving as a neutral benchmark that aggregates user feedback across all ISPs.
The platform-wide MOS reflects the average subjective quality observed across the platform and is computed over 5-minute non-overlapping time intervals. For each such interval $t$ we begin by calculating the average MOS reported by users of each $\text{ISP}_i \in \mathcal{I}$, where $\mathcal{I}$ denotes the set of all simulated providers. Denoting the average MOS for ISP$_i$ in time window $t$ as $\mu_{i,t}$, the global platform-wide MOS is defined as:
\begin{equation}
\label{formula:global-mos}
    \text{Global}_{MOS_{t}} = \frac{1}{|\mathcal{I}|} \sum_{i\in\mathcal{I}} \mu_{i,t}
\end{equation}
\autoref{formula:global-mos} ensures that each ISP contributes equally to the overall measure, avoiding biases due to volume disparities between providers. The resulting global curve--plotted in \autoref{fig:avg-5min}--captures the aggregate platform sentiment, accounting for both localized network impairments and shared platform-level phenomena such as CDN disruptions, streaming backend errors, or frontend interface latency.
However, while the global MOS trend is critical for understanding the system-wide user experience, it is insufficient for isolating ISP-specific circumstances. For this reason, we also compute the $\Delta \text{MOS}$ which isolates each ISP's relative deviation from the platform average. This is formally defined for each $ISP_i$ and time window $t$ as:
\begin{equation}
\label{formula:isp-delta-mos}
    \Delta \text{MOS\_{i,t}} = \mu_{i,t} - \texttt{Global}_{\text{MOS}_{t}}
\end{equation}

By subtracting the platform-wide MOS in \autoref{formula:isp-delta-mos}, $\Delta \text{MOS}$ allows each operator to interpret their performance relative to the shared service baseline, effectively filtering out global trends and emphasizing provider-specific quality shifts.
As demonstrated in the ISP-specific $\Delta \text{MOS}$ plots in later subsections, this dual-perspective--comparing local MOS and $\Delta \text{MOS}$-- enables operators to distinguish between platform-induced degradations and network-specific issues. The platform-wide MOS thus serves not only as a reference metric but also as a foundational element for performance accountability and anomaly attribution in multi-operator environments.

\subsection{ISP-Level $\Delta \text{MOS}$ Analysis and Outage Thresholding}
\label{sec:isp-level-mos}
\begin{figure*}
    \centering
        \includegraphics[width=0.99\linewidth]{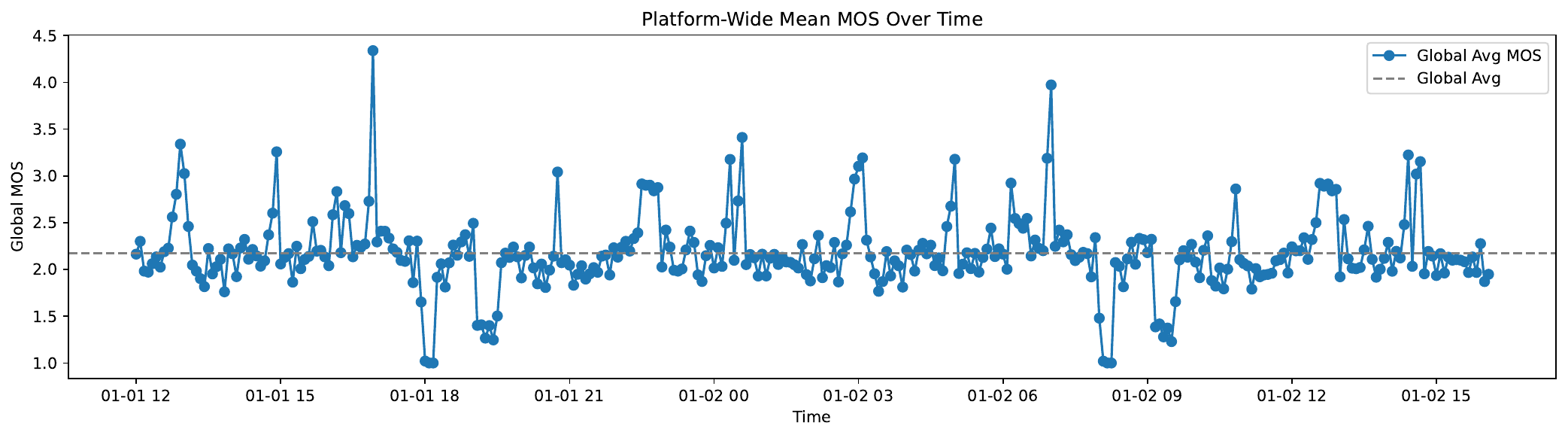}
    \caption{Platform-wide MOS over time, averaged across all ISPs in 5-minute windows. The solid line represents temporal fluctuations in user-perceived quality aggregated from subjective comment-based scores, while the dashed line indicates the overall average across the entire observation period. This global trend serves as a shared reference for evaluating relative performance of individual ISPs, as used in the $\Delta\text{MOS}$ calculations in subsequent analysis.}
\label{fig:global-mos}
\end{figure*}
\begin{figure*}
\centering
     \begin{subfigure}[b]{0.99\textwidth}
         \centering
         \includegraphics[width=\linewidth]{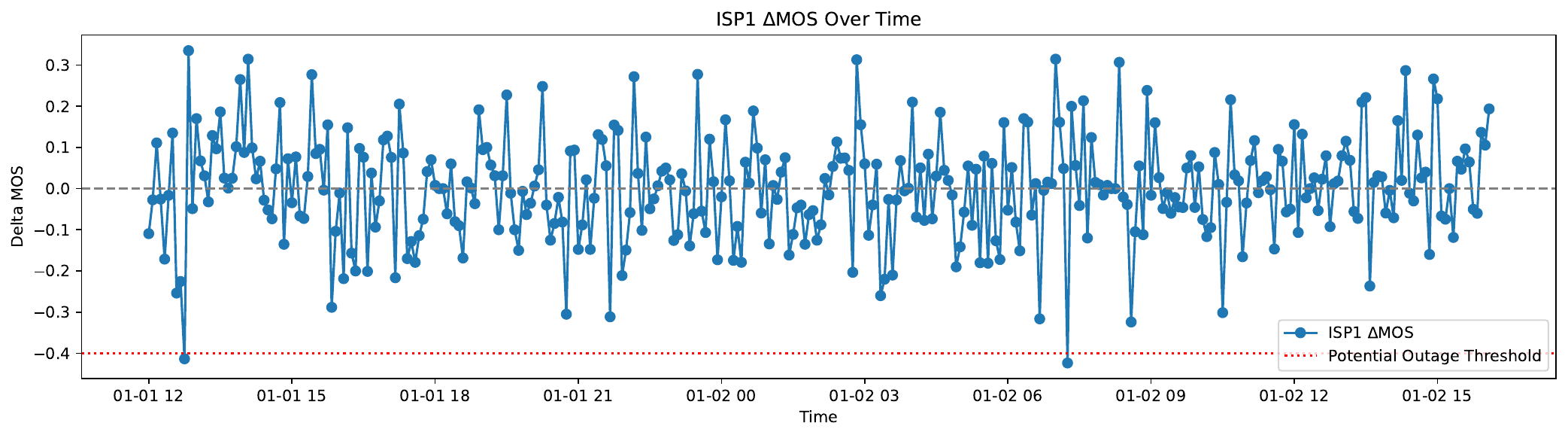}
         \caption{
             $\Delta \text{MOS}$ time series for ISP1 showing 5-minute deviations from the platform average. Values below –0.4 suggest potential service degradation.
        }
         \label{fig:isp-mos-ISP1}
     \end{subfigure}
     \hfill
     \begin{subfigure}[b]{0.99\textwidth}
         \centering
         \includegraphics[width=\linewidth]{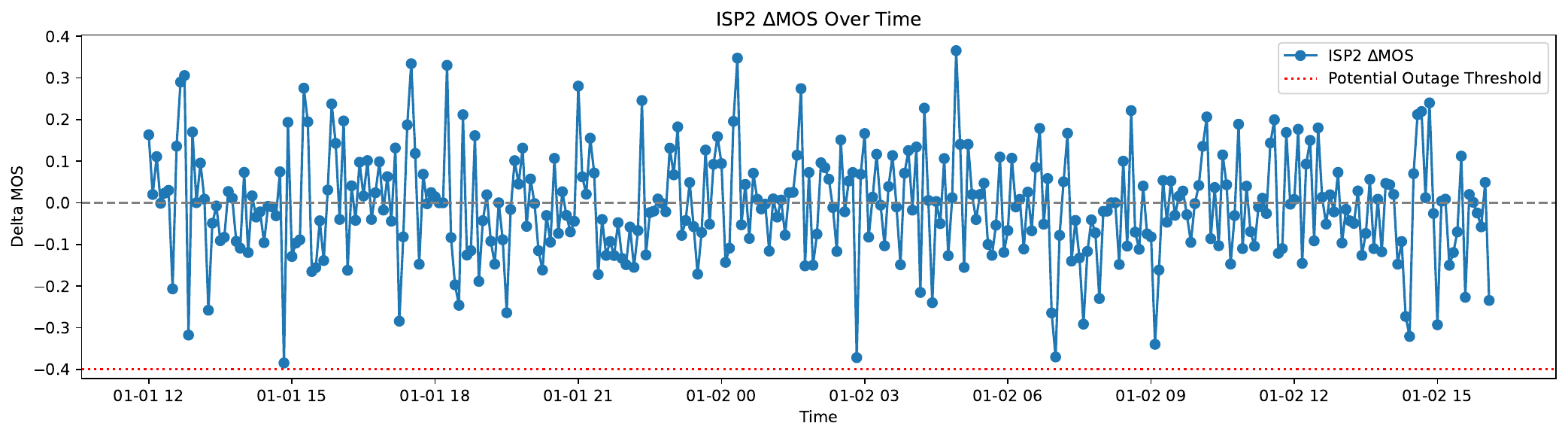}
        \caption{
            $\Delta \text{MOS}$ time series for ISP2 across 5-minute windows. Deviations from the 0 baseline show relative performance, with values below –0.4 indicating possible quality issues.
            }
        \label{fig:isp-mos-ISP2}
     \end{subfigure}
          \begin{subfigure}[b]{0.99\textwidth}
         \centering
         \includegraphics[width=\linewidth]{Pics/isp-mos-ISP2.pdf}
        \caption{$\Delta \text{MOS}$ time series for ISP3 in 5-minute intervals, highlighting quality deviations from the platform average. Drops below –0.4 suggest potential service degradation.}
        \label{fig:isp-mos-ISP3}
     \end{subfigure}
        \caption{$\Delta \text{MOS}$ time series for ISP3 in 5-minute intervals, highlighting quality deviations from the platform average. Drops below –0.4 suggest potential service degradation.}
\end{figure*}

While the platform-wide MOS offers a useful macro-level indicator of perceived service quality across the entire system, it does not account for localized performance disparities between individual ISPs. To enable provider-specific diagnostics, we introduce the concept of delta MOS ($\Delta \text{MOS}$)--a per-ISP metric that quantifies the deviation between each ISP's perceived service quality and the global platform average.
Formally, for each provider $i$ and time window $t$, let $\mu_{i,t}$ represent the provider's average MOS, and let $\text{Global}_{\text{MOS}_{t}}$ denote the platform-wide average across all ISPs in the same interval. The delta MOS is then defined in \autoref{formula:isp-delta-mos}
This formulation isolates relative quality performance: a $\Delta \text{MOS}$ close to zero indicates that the ISP is performing near the platform average, while positive or negative values respectively indicate better- or worse-than-average performance.

The evolution of $\text{Global}_{\text{MOS}_{t}}$ across time is depicted in \autoref{fig:global-mos}. This global reference curve, computed by averaging the per-ISP MOS scores within each 5-minute interval, captures the collective sentiment of all users on the platform. As such, it reflects the influence of both network and platform-side conditions and provides essential context for interpreting $\Delta \text{MOS}$ trends.
To visualize how each provider deviates from this shared baseline, we generate individual $\delta$ MOS time series plots for each ISP, shown in \autoref{fig:isp-mos-ISP1}, \autoref{fig:isp-mos-ISP2}, and \autoref{fig:isp-mos-ISP3}. These figures trace each ISP's relative quality trajectory over time, helping operators assess stability, identify anomalies, and compare service consistency.

For interpretability and threshold-based monitoring, each ISP-specific diagram includes:
\begin{enumerate}
    \item A dashed gray line at $\Delta \text{MOS}$ = 0, representing alignment with the platform average
    \item A dotted red line at $\Delta \text{MOS}$ = –0.4, marking a configurable potential outage threshold
\end{enumerate}

This threshold is not fixed by the system but is instead meant to be operator-defined, depending on business goals or SLA tolerance. A drop below this level signals a potentially significant service degradation unique to that ISP. Importantly, this threshold is independent of the global MOS trend. For instance, during periods of relatively stable global MOS (\textit{e.g.,} $\text{Global}_{\text{MOS}} \approx 2.5$), a significant negative deviation for a single provider may still indicate localized issues requiring operator intervention.
The $\Delta \text{MOS}$ framework thus enables a dual-layer perspective: it empowers operators to assess how their service compares to overall platform conditions, while filtering out platform-wide effects that could otherwise obscure local anomalies. This model supports precise root-cause attribution and provides a scalable foundation for real-time QoE monitoring, anomaly detection, and SLA enforcement.

\subsection{Simulated Outage Detection}
\begin{figure*}
    \centering
        \includegraphics[width=0.99\linewidth]{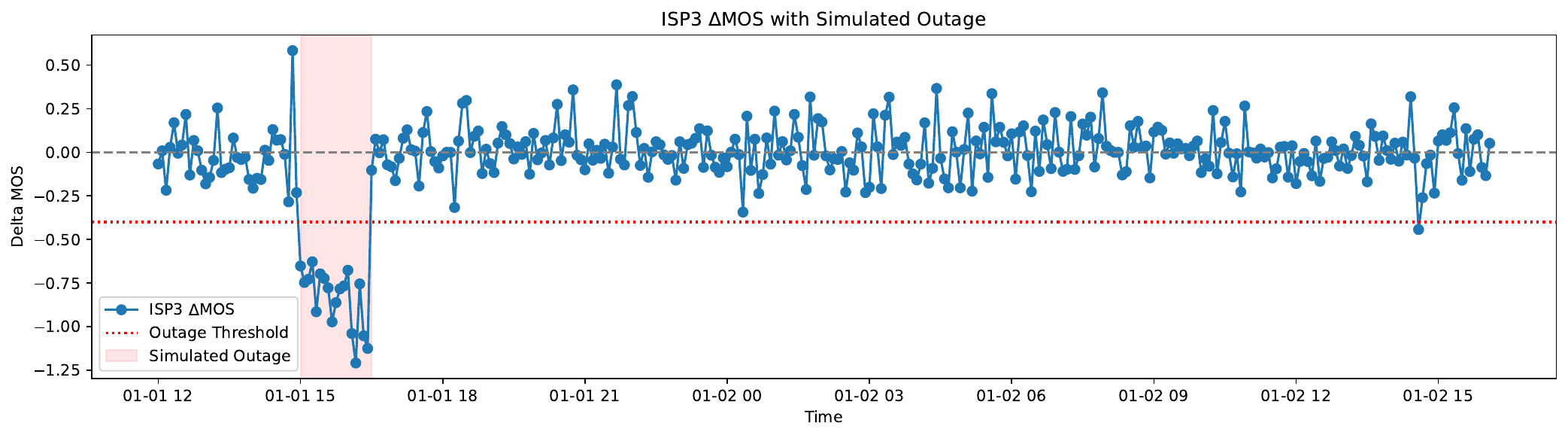}
    \caption{
    $\Delta \text{MOS}$ time series for ISP3 with a simulated outage injected between 15:00 and 16:30 on January 1, 2024. During this window, all MOS scores were manually set to 1.0 to emulate severe user dissatisfaction. The resulting  $\Delta \text{MOS}$ values fall sharply below the platform average and cross the –0.4 outage threshold (red dotted line). The red shaded region highlights the duration of the simulated degradation, demonstrating the model's ability to detect and isolate localized QoE disruptions using comment-based subjective scoring.
    }
\label{fig:isp-mos-outage-isp3}
\end{figure*}

To evaluate the responsiveness and interpretability of the $\Delta \text{MOS}$ metric in identifying service degradation, we conduct a controlled simulation of a provider-specific outage. This experiment serves as a proof-of-concept for how subjective platform-derived feedback can reveal localized anomalies in real time.
We inject a synthetic outage into the dataset by manually degrading the MOS for ISP3 within a fixed temporal interval. Specifically, for all user comments associated with ISP3 between 15:00 and 16:30 on January 1, 2024, the MOS value is forcibly set to 1.0, representing the lowest point on the QoE scale and reflecting severe user dissatisfaction. This manipulation emulates a period of significant network degradation from the end-user's perspective.

Formally, the synthetic outage is applied using the following mask:
\begin{equation}
\label{formula:outage-simulation}
    \forall i \in \text{ISP3}, \text{if }t \in [\text{15:00},\text{16:30}] \Longrightarrow \text{MOS}_i = \text{1.0}
\end{equation}
Following \autoref{formula:outage-simulation} manipulation, the data is re-aggregated at the 5-minute window level, and the per-ISP MOS scores are recalculated. The new delta MOS values are derived by subtracting the recomputed platform-wide MOS from the adjusted ISP-level averages, as described in \autoref{sec:isp-level-mos}.
The impact of this simulated outage is visualized in  \autoref{fig:isp-mos-outage-isp3}, which shows a sharp and sustained drop in ISP3's $\Delta \text{MOS}$ during the affected interval. The plot clearly reflects a negative deviation well below the –0.4 outage threshold, reinforcing the utility of $\Delta \text{MOS}$ as a reliable signal for early detection of network-related quality problems. The red-shaded vertical band in the figure marks the simulated outage window, aligning precisely with the region where ISP3's $\Delta \text{MOS}$ sharply declines.

This experiment demonstrates that even in the absence of objective, network-side measurements, subjective platform-derived feedback--when aggregated, filtered, and structured--can serve as a high-resolution QoE monitoring signal. Network operators receiving regular platform-side MOS summaries (\textit{e.g.,} per 5-minute window) can detect these anomalies, correlate them with ITU-P1203 MOS (Ref.\cite{mine-ieee}), and localize issues with minimal latency.
This example illustrates a broader capability: combining subjective QoE inference from user comments with real-time ISP attribution enables scalable anomaly detection that aligns with user experience, not just infrastructure telemetry.
\subsection{Operator Reporting Model and Interoperability}
The ultimate goal of platform-side subjective QoE analysis is to deliver actionable insights to network operators in a manner that is interpretable, timely, and directly aligned with user experience. To that end, the platform is designed to report two primary metrics to each operator for every monitoring interval (\textit{e.g.,} every 5 minutes):
\begin{enumerate}
    \item Local Operator MOS ($\mu_{i,t}$ -- the average Mean Opinion Score derived from filtered user comments attributed to operator $i$ in time window $t$
    \item Platform-Wide MOS ($\text{Global}_{\text{MOS}_t}$) -- the overall platform-wide average MOS for the same time window, computed across all ISPs.
\end{enumerate}
These two values form the basis for comparative analysis and performance accountability. From the operator's perspective, this information enables the following capabilities:
\begin{itemize}
    \item \textbf{Trend Monitoring:} By tracking their own $\mu_{i,t}$ over time, operators can observe fluctuations in user sentiment and correlate these with internal metrics such as packet loss, delay, or jitter.
    \item \textbf{Delta Interpretation:} By comparing $\mu_{i,t}$ to $\text{Global}_{MOS_t}$, operators can assess whether drops in perceived quality are localized to their infrastructure or part of a platform-wide issue affecting all providers.
    \item \textbf{QoE Validation:} Operators can juxtapose subjective MOS with objective QoE metrics computed from network telemetry or ITU-T P.1203-based models (see \cite{mine-ieee}). This triangulation allows for the identification of discrepancies, such as cases where network KPIs appear healthy but users report poor experiences--suggesting potential issues at the application layer or CDN routing.
    \item \textbf{Threshold-Based Alerts:} Using pre-configured thresholds (\textit{e.g.,} $\delta$ MOS < –0.4), operators can be alerted to periods of underperformance without relying on user complaints or internal diagnostics alone.
\end{itemize}

Importantly, the platform does not expose individual user-level data or unfiltered comment content. Instead, it aggregates and anonymizes the feedback into time-windowed summaries that preserve privacy while maximizing diagnostic value.
This reporting model provides a bridge between subjective user perception and operator-level service metrics, enabling a richer, dual-perspective understanding of QoE. It also supports real-time

\section{Conclusion and Future Work}
\label{conclusion}
\subsection{Conclusion}

This work introduced a hybrid QoE monitoring framework that combines operator-side objective estimation with platform-side subjective feedback. On the network side, we used ITU-T P.1203-aligned models to estimate MOS based solely on network parameters--delay, jitter, and throughput--without requiring access to video content or user device data. On the platform side, we designed a pipeline that filters live user comments for QoE relevance and employs the LLM to assign subjective MOS scores, offering a real-time view of user sentiment.

By aggregating subjective scores over 5-minute windows and associating them with simulated ISPs, we enabled comparative analysis using a relative metric, $\Delta\text{MOS}$, which highlights deviations from a platform-wide baseline. Visualizations and a simulated outage demonstrated the framework's ability to detect localized QoE degradations using only comment-derived insights.

The proposed system is scalable, privacy-preserving, and independent of infrastructure constraints. By providing operators with both subjective MOS scores and platform-wide benchmarks, it bridges the gap between technical performance and user experience--empowering more informed and user-centric operational decisions.

\subsection{Future Work}
Several directions may further extend the impact of this framework:
\begin{itemize}
\item LLM fine-tuning on QoE-labeled datasets to improve domain consistency and scoring robustness.
\item Integration with real operator telemetry for deeper alignment between subjective and objective metrics.
\item Multilingual and cross-platform comment processing to generalize beyond English Twitch streams.
\item Anomaly detection Assistant using $\Delta\text{MOS}$ trends and contextual metadata to identify and classify network disruptions automatically.
\item Root-cause correlation by incorporating application-side, CDN-level, or regional infrastructure indicators.
\end{itemize}
Together, these extensions aim to evolve the framework into a real-time QoE insight platform capable of assisting both operators and service providers in delivering more reliable and satisfying user experiences.

\bibliographystyle{IEEEtran}
\bibliography{References.bib}

\ignore{
\begin{IEEEbiography}[{\includegraphics[width=1in,height=1.25in,clip,keepaspectratio]{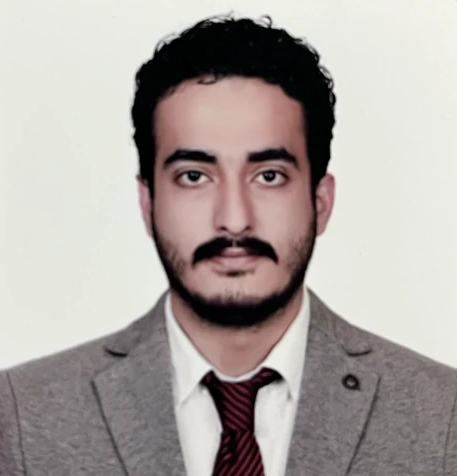}}]{Parsa Hassani Shariat Panahi }

received his B.Sc. degree in Computer Engineering from Azad University South Tehran Branch, Tehran, Iran, and the M.Sc. degree in Computer Engineering - Computer Networks from Iran University of Science and Technology, Tehran, Iran. His research interests include cellular networks, QoE assessment, telecommunication networks, wireless communication, and machine learning. He can be reached at parsa\_hassani@comp.iust.ac.ir.
\end{IEEEbiography}

\begin{IEEEbiography}[{\includegraphics[width=1in,height=1.25in,clip,keepaspectratio]{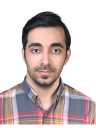}}]{Amir Hossein Jalilvand } 

received his B.Sc. degree in Computer Engineering from Bu-Ali Sina University, Hamadan, Iran, and the M.Sc. degree in Computer Engineering - Computer Architecture from Iran University of Science and Technology, Tehran, Iran. He is currently pursuing his Ph.D. in Computer Engineering. His research interests include cellular networks, stochastic and unary computing, computer architecture, fuzzy logic, and machine learning. Mr. Jalilvand has authored several publications in these fields. He can be reached at jalilvand\_a@comp.iust.ac.ir.
\end{IEEEbiography}
}

\end{document}